\documentclass[english,conference]{IEEEtran}                                                          
\IEEEoverridecommandlockouts                              
\usepackage[paper=letterpaper,top=1in,bottom=0.75in,right=0.75in,left=0.75in]{geometry}

\usepackage{amsmath}
\usepackage{amssymb}
\usepackage{graphicx}

\usepackage{babel}
\usepackage{subfigure}

\usepackage[export]{adjustbox}
\usepackage{multicol}
\makeatletter
\let\NAT@parse\undefined
\makeatother

\providecommand{\tabularnewline}{\\}

\usepackage{cite}
\newtheorem{assumption}{Assumption}\newtheorem{remark}{Remark}\newtheorem{definition}{Definition}\newtheorem{lemma}{Lemma}\newtheorem{corollary}{Corollary}
\newtheorem{innercustomthm}{Theorem}\newenvironment{customthm}[1]{\renewcommand\theinnercustomthm{#1}\innercustomthm}{\endinnercustomthm}

\begin{document}
\title{\LARGE \bf Robust Learning of Nonlinear Dynamical Systems with Safety and Stability Properties}
\author{Iman Salehi, Ghananeel Rotithor, Ashwin P. Dani
\thanks{
\newline Iman Salehi, Ghananeel Rotithor, and Ashwin P. Dani are with the Department of Electrical and Computer Engineering at University of Connecticut, Storrs, CT 06269. Email: \tt\small \{iman.salehi; ghananeel.rotithor; ashwin.dani\}@uconn.edu}}
\maketitle
\begin{abstract}
The paper presents a robust parameter learning methodology for identification of nonlinear dynamical system from data while satisfying safety and stability constraints in the context of learning from demonstration (LfD) methods. Extreme Learning Machines (ELM) is used to approximate the system model, whose parameters are learned subject to the safety and stability constraints obtained using zeroing barrier and Lyapunov-based stability analysis in the presence of model uncertainties and external disturbances. A constrained Quadratic Program (QP) is developed, which accounts for the ELM function reconstruction error, to estimate the ELM parameters. Furthermore, a robustness lemma is presented, which proves that the learned system model guarantees safety and stability in the presence of disturbances. The method is tested in simulations. Trajectory reconstruction accuracy of the method is compared against state-of-the-art LfD methods using swept error area (SEA) metric. Robustness of the learned model is tested by conducting Monte Carlo tests. The proposed method is implemented on a Baxter robot for a pick-and-place task where the robot is constrained to an ellipsoidal safety region. \end{abstract}

\section{Introduction}

In many robotics and control engineering applications, it is required to keep robot's motion within a confined set, e.g., manufacturing robotics \cite{peshkin2001cobot,xu2019robot,Zhong2021}, medical robotics \cite{Su2021} and space robotics. Imitation learning is commonly used for transferring skills to robots \cite{liu2020skill}. Learning system dynamics model using state action pair data which can be used to produce desired motion paths for robots or autonomous systems is one of the methods used for imitation learning \cite{ijspeert2002movement,schaal2003computational}. While classical system identification methods enable us to develop mathematical models of a nonlinear dynamical system from observed input-output data, identification of complex nonlinear systems requires some prior knowledge about the structure of the underlying nonlinearity \cite{pillonetto2014kernel}. On the other hand statistical methods such as Gaussian Mixture Models (GMM), Gaussian Processes (GPs), Hidden Markov Models (HMMs), Extreme learning machines (ELMs) do not require prior knowledge of the model structure. However, preserving the properties of nonlinear dynamical systems such as stability, safety and robustness to disturbances, which provides resiliency property, present challenges in developing data-driven model learning methods \cite{khosravi2020nonlinear,umlauft2017learning}.
Commonly used methods for generating robot motions using model learning include dynamic movement primitives (DMP) \cite{ijspeert2013dynamical,vijayakumar2005incremental,yang2018robot}, learning dynamical system models subject to Lyapunov stability constraints \cite{lemme2015open,neumann2013optimizing,khansari2017learning}, contraction constraints \cite{chaandar2019learning,chaandar2018learning,singh2017robust}, Riemannian motion policies \cite{ratliff2018riemannian}, control Lyapunov functions (CLF) \cite{khansari2014learning}.  However, these methods do not encode prior knowledge of safety in data-driven model learning.

The concept of safety is centered around the idea of constraining the solutions of the system to a prescribed set, by ensuring forward invariance of the set with respect to the system model. Barrier function (BF) is a commonly used approach to certify the forward invariance of a closed set with respect to a system model. Control barrier functions (CBFs) are used for the synthesis of safety-critical controllers via Quadratic Programming (QP) \cite{nguyen2016exponential,berkenkamp2016safe,ames2016control}. Since CBF is a Lyapunov-like control design, CLF and CBF are merged to synthesize stable and safe controllers by solving a QP for cyber-physical systems in \cite{romdlony2016stabilization,jankovic2017combining}. Controllers that are synthesized using CLFs and CBFs method require extensive tuning in the presence of disturbances and model uncertainties. In \cite{taylor2020adaptive}, an adaptive CBF (aCBF) is proposed that ensures forward invariance of a closed set with respect to a nonlinear control-affine system with parameter uncertainties. The controller and the parameter update laws are computed by solving an optimization problem. In \cite{Lopez2021robust}, the aCBF method is merged with a new adaptive data-driven safety paradigm for systems nominally contracting in closed-loop. These methods mainly focus on controller synthesis problem with a known dynamic system model. In \cite{dean2019safely} simultaneous linear system model identification and linear quadratic regulator (LQR) synthesis is achieved by safely exploring the state space given state and control input constraints. 

Barrier Lyapunov function is another method that is used to control nonlinear systems with output and state constraints \cite{ngo2005integrator,tee2011control,salehi2020safe}. An admittance controller is designed in \cite{he2020admittance} to ensure the robot's end-effector motion complies with the human-robot working space constraints. In developing the controller, a learning method based on the radial basis function neural network is utilized to overcome the dynamic uncertainties and improve tracking performance. Within the same spirit, an adaptive fuzzy neural network controller for multiple coordinated robots with unknown dynamics and environments carrying an object using impedance learning is introduced in \cite{kong2019adaptive}. In designing the controller, the integral barrier Lyapunov function is utilized to avoid state and output constraints imposed from the robots movements in a confined space at a limited velocity.

The model learning problem used in dynamic system learning for imitation learning of learning from demonstration (LfD) methods is closely related to the system identification methods using machine learning for nonlinear dynamical systems \cite{pillonetto2014kernel}. In our prior work \cite{salehi2019active, salehi2021dynamical}, an active learning method is presented to encode safety properties in data-driven model learning using reciprocal BF constraint. Although reciprocal barrier functions can guarantee safety property, they are not robust to the disturbances acting on the system \cite{xu2015robustness}. The lack of robustness can lead to violations of safety property in the sense of robots operating in a desired safe region will not be able to re-enter if they leave the safe region due to disturbances. Zeroing BFs, on the other hand, are robust to external disturbances \cite{xu2015robustness}.

The main contribution of this paper is to develop a nonlinear dynamical system model learning method with stability and safety properties embedded in the model learning such that the resulting dynamics is robust to external disturbances with respect to forward invariant set. Safety is encoded using zeroing BFs and stability is encoded using Lyapunov function (LF) constraints. An ELM  is used to approximate the unknown continuous nonlinear function of the dynamical system. Hence, the proposed method is called as ELM with zeroing barrier and Lyapunov functions (ELM-ZBFLF). The parameter learning problem is formulated as an optimization problem subject to zeroing BF and LF constraints. There are two main challenges in solving the formulated constrained model learning optimization problem. The constraints should be satisfied at all the points in the invariant set which leads to a semi-infinite program, and how to make the constrained optimization problem implementable in practice considering the ELM reconstruction error which cannot be measured. To overcome the first challenge, an equivalent optimization problem is derived which uses constraints evaluated at finite number of sample points inside the invariant set. Theorems are developed to show that constraints at a finite number of sample points are sufficient to ensure that the given safe set is forward invariant for the learned model and the trajectories are uniformly ultimately bounded (UUB) with respect to an equilibrium point inside the set. To overcome the second challenge, modified constraints are derived that are implementable without measuring the reconstruction error. The robustness property of the learned model is verified in the presence of additive disturbances and corresponding robustness analysis is presented. The analysis shows that when the external bounded disturbance pushes the system trajectories outside the invariant set, the system trajectories of the learned model asymptotically reaches the invariant region due to robustness properties of the zeroing BF. Similarly, the Lyapunov constraint provides exponential convergence to the equilibrium points without the ELM reconstruction error. The robustness analysis shows that trajectories of the learned model stays within a finite bound with respect to the equilibrium point in the presence of ELM reconstruction error and bounded disturbance.

Further, the proposed model learning method is tested on a trajectory data from a publicly available dataset. Accuracy metrics are computed using swept error area (SEA) metric for the proposed method, and are compared to the state-of-the-art algorithms for dynamic system learning methods of LfD. Robustness of the learned model is tested by adding disturbance to the learned model. Monte Carlo simulation test is conducted to report on bound violations during the trajectory reproduction using the metric computed based on the bound derived in Lemma \ref{lem:robustness}. The method's ability to learn and reproduce motions using an existing low-level robot controller from real-world data is illustrated using Baxter robot. A confined shelf is chosen as a limited working space for the robot to perform a pick-and-place task while following a motion trajectory that avoids an obstacle and remains inside the workspace without exiting its the boundaries.

\section{Preliminaries}

\subsection{Barrier Function}\label{subsec:EBF}

Consider an unknown locally Lipschitz continuous nonlinear dynamical system of the form 
\begin{align}
    \dot{x}(t)=f(x(t)),\label{eq:orig_dyn}
\end{align} where $f\!:\mathbb{R}^n \rightarrow \mathbb{R}^n$ is a continuous nonlinear function and $x(t)\in \mathcal{S}\subseteq \mathbb{R}^n$ is the state of the system. The system in (\ref{eq:orig_dyn}) is assumed to be \textit{forward complete}, that is, for any initial condition $x(0)\!:=\!x(t_0)\in \mathbb{R}^n$, there exists a maximum time interval denoted by $I(x(0))=[t_0,\infty)$ such that $x(t)$ is the unique solution to the corresponding differential equation \cite{ames2016control}. Throughout this paper, for ease of notation, the dependency of $x(t)$ is abbreviated by $x$, unless necessary for clarity.

A closed set $\mathcal{S}\in \mathbb{R}^n$, its interior, and its boundary are defined as 
\begin{align}
\mathcal{S} & =\{x\in\mathbb{R}^{n}:\ h(x)\geq0\},\label{eq:S-1}\\
\partial\mathcal{S} & =\{x\in\mathbb{R}^{n}:\ h(x)=0\},\label{eq:S-2}\\
\mathrm{Int}(\mathcal{S}) & =\{x\in\mathbb{R}^{n}:\ h(x)>0\},\label{eq:S-3}
\end{align}
where $h\ :\ \mathbb{R}^{n}\rightarrow\mathbb{R}$ is a continuously differentiable function. 

\begin{assumption}\label{assump_1}
The function $f:\mathbb{R}^{n}\rightarrow\mathbb{R}^{n}$ is locally Lipschitz continuous with Lipschitz constant $L_{f}$ and bounded in $\mathcal{S}$ with $\Vert f(\cdot)\Vert\leq\bar{f}$, where $\bar{f}$ is a positive scalar. 
\end{assumption}

\begin{definition}\label{def:1}\cite[Definition 3]{ames2016control}
Given the continuous system (\ref{eq:orig_dyn}), the set $\mathcal{S}$ defined by (\ref{eq:S-1})-(\ref{eq:S-3}), a continuously differentiable function $h\!:\mathbb{R}^n \rightarrow \mathbb{R}$ is said to be a zeroing BF, if there exist a locally Lipschitz extended class $\mathcal{K}$ function $\eta$ and a set $\mathcal{D}$ with $\mathcal{S} \subseteq \mathcal{D} \subset \mathbb{R}^n$ such that for all $x\in \mathcal{D}$
\begin{align}
  \nabla h(x)^T f(x)& \geq -\eta(h(x)), \label{eq:ZBF-cond}
\end{align}
\end{definition}
where the operator $\nabla: C^1(\mathbb{R}^n) \rightarrow \mathbb{R}^n $ is defined as the gradient $\frac{\partial }{\partial x}$ of a scalar-valued differentiable function with respect to $x$.

Using Nagumo's theorem, \cite{blanchini2008set}, 
it can be shown that with a zeroing BF defined in Definition \ref{def:1}, for any $x\in \partial\mathcal{S}$, $\nabla h(x)^T f(x)\geq -\eta(0)=0$, thus the set $\mathcal{S}$ is \textit{forward invariant} with respect to (\ref{eq:orig_dyn}), which means for every initial condition $x(t_0)\in \mathcal{S}$, $x(x(t_0),t)\in \mathcal{S}$, $\forall \in I(x(0))$ \cite{ames2016control}.

\begin{assumption}\label{assump_2}
The real-valued $h(x)$ is a $C^2$ differentiable function with Lipschitz constant $L_h$ such that $\Vert\nabla h(x)\Vert\leq L_h$, and $\nabla h(x)$ is globally Lipschitz continuous with Lipschitz constant $L_{\partial h}$.
\end{assumption}

\begin{definition}\label{def:2}
A closed and forward invariant set $\mathcal{S}\subseteq \mathbb{R}^n$ is said to be locally asymptotically stable for a forward complete system (\ref{eq:orig_dyn}), if there exist an open set $\mathcal{D}$ containing $\mathcal{S}$ and a class $\mathcal{KL}$ function $\zeta$ such that 
\begin{align}
    \underset{s \in \mathcal{S}}{\mathrm{inf}} \Vert x-s \Vert \leq \zeta\left(\underset{s \in \mathcal{S}}{\mathrm{inf}}\Vert x(0)-s \Vert,t\right), 
\end{align}
for any $x(0)\in \mathcal{D}$; and  $\underset{s \in \mathcal{S}}{\mathrm{inf}} \Vert x-s \Vert$ denotes the distance from $x$ to set $\mathcal{S}$.
\end{definition}

\begin{remark}
Asymptotic stability of set $\mathcal{S}$ implies invariance of $\mathcal{S}$, since when $x(0)\in \mathcal{S}$ implies $\zeta\left(\underset{s \in \mathcal{S}}{\mathrm{inf}}\Vert x(0)-s \Vert\right) = 0$, which also implies $ \underset{s \in \mathcal{S}}{\mathrm{inf}} \Vert x-s \Vert =0$, thus $x(t)\in \mathcal{S}$ for all time \cite{xu2015robustness}. 
\end{remark}

\subsection{Extreme Learning Machines}

ELM is a learning algorithm that uses least squares approach to estimate the parameters of the NN. According to the ELM theory \cite{huang2006universal}, input weights and all the hidden node parameters are randomly assigned.
The choice of ELM over traditional gradient-based backpropagation training methods for NN is primarily based on faster learning speed of ELM. This choice also avoids potential issues with the gradient-based methods, such as local minima, vanishing gradients and improper learning rate \cite{huang2006universal}. Consider the following multilayer perceptron with $n_{h}$ hidden nodes 
\begin{align}
    y=\sum_{i}^{n_{h}}G_{i}(x,U_{i},a_{i},b_{i})\cdot W_{i}^{T}+\mathfrak{b}_o,
\end{align}where $y\in\mathbb{R}^{n}$, $G_{i}(\cdot)$ is a scalar, which denotes the $i^{\mathrm{th}}$ hidden node activation function whose range depends on the choice of the nonlinear function, $U_{i}\in\mathbb{R}^{n}$ is the input weight vector, \textbf{$a_{i},b_{i}\in\mathbb{R}$ } are slope and bias corresponding to the $i^{\mathrm{th}}$ hidden neuron, $W_{i}^{T}\in\mathbb{R}^{n}$ is the output weight vector, and $\mathfrak{b}_o\in \mathbb{R}^n$ is the bias vector corresponding to the output layer. The ELM is created by randomly initializing the input weight matrix $U\in\mathbb{R}^{n\times n_{h}}$, where $U_i$ are the columns of $U$, the slopes (usually are set to one) $a\in\mathbb{R}^{n_{h}}$, and the biases $b=[b_1, b_2,...b_{n_h}]^T\in\mathbb{R}^{n_{h}}$.

\section{Problem Formulation and Solution Approach}\label{sec: Problem-Formulation}
\subsection{Constrained ELM Learning Problem}
Consider a set of $N$ trajectories consisting of both the states $\{x(t)\}_{t=0}^{t=T_{n}}$, the solutions to the system (\ref{eq:orig_dyn}), and state derivatives $\{\dot{x}(t)\}_{t=0}^{t=T_{n}}$. If the state derivatives are not available, they can be computed using numerical differentiation methods such as the finite difference method. Assuming that the nonlinear function $f(\cdot)$ in (\ref{eq:orig_dyn}) is deterministic, the system in (\ref{eq:orig_dyn}) can be approximated by a NN given by
\begin{align}
\dot{x}(t) & =f(x(t))=W^{T}g+\epsilon(x(t)),\label{eq:compact_NN}
\end{align}
where $W\in\mathbb{R}^{\left(n_{h}+1\right)\times n}$ is the bounded constant output layer weight matrix, $g=[\sigma(q\!\cdot\! s(t)),1]\!\in\! \mathbb{R}^{\left(n_{h}+1\right)}$, $\sigma(\cdot)\in \mathbb{R}^{n_h}$ is the vector sigmoid activation function, $n_{h}$ is the number of neurons in the hidden layer of the NN, and $\epsilon(x)\in \mathbb{R}^n$ is the function reconstruction error. The NN used in this paper is an ELM network. The vector sigmoid activation function is given by
\begin{align}
    \sigma\left(\!q\!\cdot\!s\left(t\right)\!\right)\!=\!\left[\frac{1}{1\!+\!e^{-\left(q\cdot s\left(t\right)\right)_{1}}},\cdots,\frac{1}{1\!+\!e^{-\left(q\cdot s\left(t\right)\right)_{n_{h}}}}\right]^{T},
 \label{eq:sigmoid}
\end{align}
where $q=[P,\ b_{p}]\in\mathbb{R}^{n_{h}\times n+1}$,  $P\!=\!\mathrm{diag}(a_{p})\cdot U^{T}\in\mathbb{R}^{n_{h}\times n}$, $a_p\in\mathbb{R}^{n_h}$ and  $b_p\in \mathbb{R}^{n_h}$ are the internal slopes and biases vectors respectively,  $U\in\mathbb{R}^{n\times n_{h}}$ is the bounded input layer weight matrix, and $s\left(t\right)=\left[x\left(t\right)^{T},1\right]^{T}\in\mathbb{R}^{n+1}$ is the input vector. Other activation functions such as radial basis functions, tangent sigmoid can also be used.

\begin{remark}\label{remark_1}
The nonlinear function $f(x)$ is within real and positive constant $\bar{\epsilon}$ of the ELM range if there exist a finite number of hidden neurons $n_h$ and constant weights so that $\forall x(t)\in \mathcal{S}$  the approximation in (\ref{eq:compact_NN}) holds with $\Vert \epsilon(x) \Vert \leq \bar{\epsilon}$. The boundedness of the reconstruction error $\epsilon(x)$ follows from the \textit{Universal Approximation Property} of ELM and the fact that for any randomly generated activation function the reconstruction error goes to zero \cite{huang2006universal}, i.e.,$\underset{n_{h}\rightarrow\infty}{\mathrm{lim}}\epsilon(x)=0$.
\end{remark}
\begin{assumption}\label{assump_3}
The ideal weights of the ELM are bounded by known positive constants, i.e., $\Vert W \Vert_{F} \leq \bar{W}$, $\Vert U \Vert_{F} \leq \bar{U}$, where $\Vert \cdot \Vert_F$ is the Frobenius norm \cite{huang2006universal}. Also, it is assumed that  $\Vert \epsilon^{\prime}(x) \Vert \leq \bar{\epsilon}^{\prime}$, where prime denotes the derivative with respect to $x$.
\end{assumption}

\begin{remark}\label{remark_2}
The sigmoid function $\sigma_i(\cdot)\in [0,1]$ and hence its derivative $\sigma_i(\cdot)(1-\sigma_i(\cdot))$ has upper and lower bounds given by $0\leq\sigma_i(\cdot)(1-\sigma_i(\cdot))\leq0.25$ $\ \forall i = 1,\cdots n_h$. Thus, $\Vert \sigma(\cdot) \Vert \leq \sqrt{n_h}$ and $\Vert \sigma(\cdot)(1-\sigma(\cdot))\Vert 
\leq 0.25\sqrt{n_h}$. Using Assumption \ref{assump_3}, Assumption \ref{assump_1} is still valid for the ELM parameterization of $f(x)$ \cite{lewis2002neuro}.
\end{remark}

\paragraph*{Problem Statement} Given a set of trajectory data $\{x(t)\}_{t=0}^{t=T_n}$ and corresponding state derivatives $\{\dot{x}(t)\}_{t=0}^{t=T_n}$, the problem of learning the nonlinear function $f(\cdot)$ in (\ref{eq:compact_NN}), approximated using ELM, is addressed such that the solution trajectories of the system in (\ref{eq:compact_NN}) satisfy safety property defined as a forward invariance of a set constructed using zeroing barrier functions, and a stability property defined with respect to a specified equilibrium point inside the invariant region derived using a Lyapunov stability analysis. 

\subsection{Encoding Safety Constraints}
To achieve the safety property, the zeroing BF formulation is presented in this subsection, which is used to derive safety constraints used in the constrained parameter learning of the ELM.

\subsubsection{ELM Parameter Learning Problem with Zeroing BF}

As mentioned in Subsection \ref{subsec:EBF}, zeroing BF defines a forward invariant set such that solutions of the nonlinear dynamical system that start in that set remain in that set for all time. An example of a zeroing barrier function $h(x)$ in $\mathbb{R}^2$ is given below. 

Example of a zeroing BF set in $\mathbb{R}^2$: Equation of an ellipse is used as an example of zeroing BF in $\mathbb{R}^2$. Consider a closed and connected set $\mathcal{X}_0 \subset \mathcal{X}\subset \mathbb{R}^2$, which is defined according to (\ref{eq:S-1}) in Section \ref{subsec:EBF}. The function $h(x)\!:\!\mathbb{R}^2 \rightarrow \mathbb{R}^+$, where $x = [x_1, x_2]^T$ is selected as follows
\begin{align}
h(x) = & \ 1-\frac{((x_{1}-x_{1g})\cos\alpha+(x_{2}-x_{2g})\sin\alpha)^{2}}{\iota_{1}^{2}} \nonumber \\
& -\frac{((x_{1}-x_{1g})\sin\alpha+((x_{2}-x_{2g})\cos\alpha)^{2}}{\iota_{2}^{2}},
\label{eq:ellipse}
\end{align}
where $x_{1g}$ and $x_{2g}$ are the coordinates of the center of gravity of the ellipse, $\iota_{1}$ and $\iota_{2}$ are the major and minor axes of the ellipse respectively, and $\alpha$ is the orientation of ellipse. 
\begin{remark}
Based on the application any arbitrary shape of the invariant region can be chosen for which the BF can be designed as long as it satisfies the conditions defined in (\ref{eq:S-1})-(\ref{eq:S-3}) and Assumption \ref{assump_2}. On defining the invariant region, the continuously differentiable function $h(x)$ in (\ref{eq:ellipse}) can also be constructed for the higher dimensions.  
\end{remark}

Given the system model in (\ref{eq:compact_NN}) ELM parametrization and zeroing BF constraint, the ELM parameter learning problem with zeroing BF constraints is now discussed. The constrained optimization problem for ELM training is formulated as follows:
\begin{align}
W^{*} & =\arg\:\min_{W}\:\sum_{t=0}^{T_{n}} E_D\label{eq:objFunc}\\
\mathrm{s.t.} & \quad \dot{h}(x)\geq -\gamma h(x),\quad\forall x\in\mathcal{X},\label{eq:ZBF-constr}
\end{align} where $E_D = \left[\dot{x}(t)\!-\!\hat{\dot{x}}(t)\right]^{T}\!\!\left[\dot{x}(t)\!-\!\hat{\dot{x}}(t)\right]\!+\!\mu_{W}\!\left(tr\left(W^{T}W\right)\right)$ is the sum of squared error, $\dot{x}(t)\in\mathbb{R}^{n}$ and $\hat{\dot{x}}(t)\in\mathbb{R}^{n}$ represent the target and the ELM's output, $\mu_{W}\in\mathbb{R}^{+}$ is the regularization parameter.

Using the constraint (\ref{eq:ZBF-constr}) in the ELM parameter learning optimization problem for learning the nonlinear dynamical system is challenging because the constraint (\ref{eq:ZBF-constr}) must be satisfied $\forall x\in\mathcal{X}$, which is an uncountably infinite set. In addition, the ELM reconstruction error is not measurable. To overcome the first stated challenge, an equivalent optimization problem is formulated, which uses finite number of constraints developed by evaluating (\ref{eq:ZBF-constr}) at finite number of points in $\mathcal{X}$ and utilizing Lipschitz continuity of function $h(x)$ in $x$ with Lipschitz constant $L_h$. Theorem \ref{thm:1} is developed to prove that enforcing a modified barrier constraint for a finite number of sampled points is sufficient for providing solution to the optimization problem in (\ref{eq:objFunc})-(\ref{eq:ZBF-constr}). For the second stated challenge of ELM reconstruction error, new constraint is derived which is independent of ELM function reconstruction error, and uses an upper bound to make it implementable in practice. Corollary \ref{cor:1} is presented to prove that the newly derived constraint is sufficient for the constraint in (\ref{eq:ZBF-constr}) to hold.  

Let $\mathcal{X}_{\tau}\subset\mathcal{X}$ be a discretization of the state space $\mathcal{X}$ with the closest point in $\mathcal{X}_{\tau}$ to $x\in\mathcal{X}$ denoted by $[x]_{\tau}$ such that $\Vert x-[x]_{\tau}\Vert\leq\frac{\tau}{2}$, and $\tau$ is the discretization resolution. 
\begin{customthm}{1}\label{thm:1}
Given Assumptions \ref{assump_1}-\ref{assump_3}, if the following condition holds 
\begin{align}
-\nabla h\left([x]_{\tau}\right)^T f([x]_{\tau}) - \gamma h\left([x]_{\tau}\right) & \leq-\mathcal{L}_{\dot{h}}-\gamma L_{h}\frac{\tau}{2}, \label{eq:thm_1}
\end{align} $\forall [x]_{\tau}\in \mathcal{X}_{\tau}$, where $\mathcal{L}_{\dot{h}}=\left(L_{\partial h}\left(\bar{W}\bar{g}+\bar{\epsilon}\right)+L_{h}L_{f}\right)\frac{\tau}{2}$ and $\bar{g}=\sqrt{n_{h}+1}$, then the safety zeroing barrier constraint in (\ref{eq:ZBF-constr}) is satisfied $\forall x \in \mathcal{X}$.
\end{customthm} 

\begin{IEEEproof}
Using $\dot{h}(x)=\nabla h(x)^T f(x)$, (\ref{eq:thm_1}) can be rewritten as $-\dot{h}\left([x]_{\tau}\right)-\gamma h\left([x]_{\tau}\right)\leq-\mathcal{L}_{\dot{h}}-\gamma L_{h}\frac{\tau}{2}$, $\forall[x]_{\tau}\in\mathcal{X}_{\tau}$, which implies 
\begin{align}
-\dot{h}\left([x]_{\tau}\right)\!-\!\gamma h\left([x]_{\tau}\right)\!+\!\mathcal{L}_{\dot{h}}+\gamma L_{h}\frac{\tau}{2}\leq0, &\  \forall[x]_{\tau}\in\mathcal{X}_{\tau}. \label{eq:thm_1_2}
\end{align} If $-\dot{h}(x)-\gamma h(x)$ is a lower bound on the LHS of (\ref{eq:thm_1_2}), then the zeroing barrier constraint in (\ref{eq:ZBF-constr}) will be satisfied. Consider $\mathcal{O}=-\left(\dot{h}(x)-\dot{h}\left([x]_{\tau}\right)\right)-\gamma\left(h(x)-h\left([x]_{\tau}\right)\right)$, which by using the triangle inequality,  $\mathcal{O}$ can be upper bounded as 
\begin{align}
    \mathcal{O}\leq \left\Vert\dot{h}(x)-\dot{h}\left([x]_{\tau}\right)\right\Vert+\gamma\Big\Vert\left(h(x)-h\left([x]_{\tau}\right)\right)\Big\Vert. \label{eq:upprbnd_O}
\end{align}Adding and subtracting $\nabla h\left([x]_{\tau}\right)f(x)$ to the first term on the RHS of (\ref{eq:upprbnd_O}), and using Assumptions \ref{assump_1}-\ref{assump_3} we have
\begin{align}
 & \left\Vert \left(\nabla h(x)\!-\!\nabla h\left([x]_{\tau}\right)\right)^{T} f(x)\!+\!\nabla h\left([x]_{\tau}\right)^T\left(f(x)-f\left([x]_{\tau}\right)\right)\right\Vert \nonumber\\
 & \leq\left\Vert \!\left(\nabla h(x)\!-\!\nabla h\left([x]_{\tau}\right)\right)\!\right\Vert \!\left\Vert W^{T}g\!+\!\epsilon(x)\right\Vert \!+\!\left\Vert \nabla h\left([x]_{\tau}\right)\right\Vert \nonumber\\
 & \cdot\left\Vert \left(f(x)\!-\!f([x]_{\tau})\right)\right\Vert \leq\mathcal{L}_{\dot{h}}, \label{eq:hdot_upper}
\end{align}where $\mathcal{L}_{\dot{h}}$ is an upper bound on $\left\Vert\dot{h}(x)-\dot{h}\left([x]_{\tau}\right)\right\Vert$ and $\bar{g}=\sqrt{n_h+1}$. Furthermore, by using Lipschitz continuity of $h(x)$, $\forall x \in \mathcal{X}$, and thereby mean value theorem, (\ref{eq:upprbnd_O}) can be rewritten as follows
\begin{align}
    \mathcal{O}&\leq \mathcal{L}_{\dot{h}}+\gamma \left(\left\Vert \nabla h(\xi)\right\Vert\Vert x-[x]_{\tau} \Vert\right)\leq \mathcal{L}_{\dot{h}} + \gamma L_h\frac{\tau}{2}, \label{eq:upprbnd_O_1}
\end{align}where $\xi \in (x,[x]_{\tau})$ is a point on the line segment connecting $x$ to $[x]_{\tau}$. Hence, $\mathcal{O}\leq  \mathcal{L}_{\dot{h}} \!+\! \gamma L_h\frac{\tau}{2}$, which yields 
\begin{align}
   -\dot{h}(x) \!-\!\gamma h(x) \!\leq\! -\dot{h}([x]_{\tau})\!+\!\mathcal{L}_{\dot{h}}\!+\!\gamma \left(\!L_h\frac{\tau}{2}\!-\!h([x]_{\tau})\!\right)\!, \label{eq:thm_1_3}
\end{align}$\forall x \in \mathcal{X} $ and $\forall [x]_{\tau}\in \mathcal{X}_{\tau}$. From (\ref{eq:thm_1_2}) and (\ref{eq:thm_1_3}), $-\dot{h}(x)-\gamma h(x) \leq 0$, $\forall x\in \mathcal{X}$.
\end{IEEEproof}

Using the result of Theorem \ref{thm:1}, substituting (\ref{eq:compact_NN}) in (\ref{eq:thm_1}), the constraint in  (\ref{eq:ZBF-constr}) can be reformulated as 
\begin{align}
    -\!\nabla h\left([x]_{\tau}\!\right)^T\!\!\left(W^{T}\!g\!+\!\epsilon\!\left([x]_{\tau}\right)\!\right)\!\!-\!\gamma h([x]_{\tau}\!)\!\leq\!-\!\mathcal{L}_{\dot{h}}\!-\!\gamma L_{h}\!\frac{\tau}{2}, \! \label{eq:linear_constr}
\end{align}$\forall[x]_{\tau}\in\mathcal{X}_{\tau}$. Implementing (\ref{eq:linear_constr}) requires the knowledge of the ELM reconstruction error, i.e., $\epsilon(x)$, which is not measurable, and therefore using the zeroing BF constraint in (\ref{eq:linear_constr}) along with the objective function (\ref{eq:objFunc}) is not feasible. A modified constraint which uses an upper bound on the ELM function reconstruction error is derived to obtain a practically implementable constraint. Corollary \ref{cor:1} is developed to show that the modified zeroing BF constraint in (\ref{eq:robust-constraint}) can be used instead of (\ref{eq:linear_constr}).  
\begin{corollary}\label{cor:1}
If the following constraint is satisfied for the nonlinear dynamical system in (\ref{eq:compact_NN}), 
\begin{align}
    -\nabla h\left([x]_{\tau}\right)^TW^Tg-\gamma h([x]_{\tau})&\leq-\mathcal{E},\ \forall [x]_{\tau} \in \mathcal{X}_{\tau},  \label{eq:robust-constraint}
\end{align}where $ \mathcal{E}=\left(L_{\partial h}\left(\bar{W}\bar{g}+\bar{\epsilon}\right)+L_{h}\left(L_{f}+\gamma \right)\right)\frac{\tau}{2}+L_{h}\bar{\epsilon}> 0$, then the set defined by $h(x)$ in (\ref{eq:ellipse}) is forward invariant with respect to the system trajectories.
\end{corollary}

\begin{IEEEproof}
    Substituting (\ref{eq:hdot_upper}) in (\ref{eq:linear_constr}), using the upper bound on $\nabla h\left([x]_{\tau}\right)^
T\epsilon\left([x]_{\tau}\right)$, and performing some algebraic manipulations to (\ref{eq:linear_constr}) yields $-\nabla h\left( [x]_{\tau}\right)^TW^T g-\gamma h([x]_{\tau})\leq-\mathcal{E}$, proving the result. 
\end{IEEEproof}

\subsection{Encoding Stability Constraints}
The zeroing BF yields exponential stability with respect to the set defined by $h(x)$ \cite{ames2016control}. In this subsection, additional stability constraint with respect to an equilibrium point inside the set $h(x)$ is developed using Lyapunov-based stability analysis. The following Lemma is presented to derive a uniformly ultimately bounded (UUB) stability constraint for ELM parameter learning.

\begin{lemma}\label{lem:UUB}
The nonlinear system in (\ref{eq:compact_NN}) is said to be UUB if
\begin{align}
    \left(x-x^{*}\right)^{T}W^Tg\leq -\beta(x),\label{eq:LF_constr} 
\end{align} $\forall x \in \mathcal{X}$ where $\beta(x)=\rho (x-x^*)^T(x-x^*)$, $x^*$ is the equilibrium point, and $\rho \in \mathbb{R}^+$ is a positive constant. Furthermore, the ultimate bound on the trajectories of the system with respect to its equilibrium is given by $ \underset{t\rightarrow\infty}{\lim\sup}\Vert x-x^{*}\Vert\leq\frac{\bar{\epsilon}}{\rho}$.
\end{lemma}
\begin{IEEEproof}
Consider a quadratic Lyapunov candidate function  $V(x)=\frac{1}{2}(x-x^{*})^{T}(x-x^{*})$, $\forall x\in\mathcal{X}$. Taking the time derivative of the Lyapunov function along the trajectories of (\ref{eq:compact_NN}) yields $\dot{V}(x)=(x-x^{*})^{T}\left(W^{T} g+\epsilon(x) \right)$. Using the definition of $\dot{V}(x)$, the constraint in (\ref{eq:LF_constr}) can be written as
\begin{align}
    \dot{V}(x) \leq-\rho\Vert x-x^{*}\Vert^{2}+\left(x-x^{*}\right)^{T}\epsilon(x).\label{eq:upperbnd_dLF-dt}
\end{align} After upper bounding the second term on the RHS of (\ref{eq:upperbnd_dLF-dt}) and completing the squares, (\ref{eq:upperbnd_dLF-dt}) is given by $\dot{V}(x) \leq -\rho V(x)+\frac{\bar{\epsilon}^{2}}{2\rho}$. Invoking Theorem 4.18 of \cite{Khalil2002}, the ultimate bound is given by $\underset{t\rightarrow\infty}{\lim\sup}\Vert x-x^{*}\Vert\leq\frac{\bar{\epsilon}}{\rho}$.
\end{IEEEproof}

Implementing the constraint in (\ref{eq:upperbnd_dLF-dt}) along with the objective function in (\ref{eq:objFunc}) is not feasible. Thus, an equivalent implementable constraint is derived. In Theorem \ref{thm:2}, it is proven that using a modified constrained based on Lyapunov analysis computed at a finite number of sampled points is equivalent to the Lyapunov stability constraint in (\ref{eq:LF_constr}).

\begin{customthm}{2}\label{thm:2} Given Assumption \ref{assump_3}, if the following condition holds
\begin{align}
\left([x]_{\tau}-x^{*}\right)^{T}W^Tg &\leq-\beta\left([x]_{\tau}\right)-\mathcal{L}_{V}\frac{\tau}{2},\ \forall[x]_{\tau}\in\mathcal{X}_{\tau}\label{eq:thm:2}
\end{align}where $\mathcal{L}_{V}=L_{\dot{V}}+2\rho L_{V}+\sqrt{2V\left([x]_{\tau}\right)}\bar{\epsilon}^{\prime}+\bar{\epsilon}$, $L_V$ is a Lipschitz constant of the selected Lyapunov function, 
\begin{align}
 L_{\dot{V}} = \bar{W}\bar{g} + \bar{\epsilon} + \left\Vert\xi-x^*\right\Vert\left(\frac{\bar{a}_p \sqrt{n_h}\bar{W}_{n_{h}}\bar{U}}{4} + \bar{\epsilon}^{\prime} \right), \label{eq:UpperBoundVDot}
\end{align} $\xi\in(x,[x]_{\tau})$ is a point on the line segment connecting $x$ to $[x]_{\tau}$, $\bar{a}_p=\Vert \mathrm{diag}(a_{p})\Vert_F$, and $\bar{W}_{n_{h}}$ is the upper bound of the matrix that is obtained by deleting the last row of $W$, then the Lyapunov constraint in (\ref{eq:LF_constr}) is satisfied $\forall x\in\mathcal{X}$. 
\end{customthm} 
\begin{IEEEproof}
The constraint in (\ref{eq:thm:2}) can be rewritten as 
\begin{align}
\dot{V}\left([x]_{\tau}\right)\!-\!\left([x]_{\tau}-x^{*}\right)^T\!\epsilon\left([x]_{\tau}\right)\!+\!\mathcal{L}_{V}\frac{\tau}{2}\!+\!\beta\left([x]_{\tau}\right)\leq 0, \label{eq:ModifiedLyapunovConstraint}
\end{align}for all  $[x]_{\tau}\in\mathcal{X}_{\tau}$. If $\dot{V}(x)+\beta(x)-\left(x-x^{*}\right)^T\epsilon(x)$ is an upper bound on (\ref{eq:ModifiedLyapunovConstraint}) then the Lyapunov constraint given in (\ref{eq:LF_constr}) is satisfied. Using the mean value theorem for $\dot{V}(x)$ yields
\begin{equation}
\dot{V}(x)-\dot{V}([x]_{\tau})=\nabla\dot{V}\left(\xi\right)\left(x-[x]_{\tau}\right),\ \forall \xi\in(x,[x]_{\tau}). \label{eq:MVT-1}
\end{equation} The LHS of (\ref{eq:MVT-1}) can be upper bounded by $\dot{V}(x)-\dot{V}([x]_{\tau})\leq\left\Vert \nabla \dot{V}\left(\xi\right)\right\Vert \left \Vert(x-[x]_{\tau})\right \Vert$. The upper bound on $\nabla\dot{V}\left(\xi\right) = \left[\nabla(x-x^{*})^{T}\left(W^{T}g+\epsilon(x)\right)\right]_{x=\xi}$ can be obtained as $\left\Vert \nabla\dot{V}(\xi)\right\Vert \leq L_{\dot{V}}$, which $L_{\dot{V}}$ is an upper bound given by (\ref{eq:UpperBoundVDot}) using Remarks \ref{remark_1} and \ref{remark_2}. Now, consider 
$\mathcal{O}_{v}\!=\!\left(\dot{V}(x)\!-\!\dot{V}([x]_{\tau})\right)\!+\!\left(\beta(x)\!-\!\beta([x]_{\tau})\right)\!+\!\left[\left([x]_{\tau}\!-\!x^{*}\right)^{T}\epsilon([x]_{\tau})\!-\!\left(x-x^{*}\right)^{T}\epsilon(x)\right]$, which can be upper bounded as 
\begin{align}
    \mathcal{O}_{v}\leq & \left\Vert \left(\dot{V}(x)-\dot{V}([x]_{\tau})\right)\right\Vert +2\rho\left\Vert \left(V(x)-V([x]_{\tau})\right)\right\Vert \nonumber \\
     & +\left\Vert \left[\left([x]_{\tau}-x^{*}\right)^{T}\!\epsilon([x]_{\tau})-\left(x-x^{*}\right)^{T}\!\epsilon(x)\right]\right\Vert. \label{eq:Ov_bound}
\end{align}Adding and subtracting $[x]_{\tau}^T\epsilon(x)$ to the third term on the RHS of (\ref{eq:Ov_bound}), using Assumption \ref{assump_3} and the triangle and Cauchy-Schwartz inequalities we have 
\begin{align}
\mathcal{O}_{v}\leq & \left(L_{\dot{V}}\!+\!2\rho L_{V}\!+\sqrt{2V\left([x]_{\tau}\right)}\bar{\epsilon}^{\prime}\!+\!\bar{\epsilon}\right)\frac{\tau}{2}=\mathcal{L}_{V}\frac{\tau}{2}\label{eq:Ov_bound-1}
\end{align}
Hence $\dot{V}(x)+\beta(x)-\left(x-x^{*}\right)^{T}\epsilon(x)\leq\dot{V}\left([x]_{\tau}\right)+\beta\left([x]_{\tau}\right)-\left([x]_{\tau}-x^{*}\right)^{T}\epsilon\left([x]_{\tau}\right)+\mathcal{L}_{V}\frac{\tau}{2}\leq 0$, which can be rearranged to obtain the constraint  in (\ref{eq:thm:2}).
\end{IEEEproof}
 
\subsection{Encoding Combined Safety and Stability Constraints}
The combined safety and stability properties are incorporated in the ELM parameter learning using Corollary \ref{cor:1} and Theorem \ref{thm:2}. Using the zeroing barrier constraint in (\ref{eq:robust-constraint}) and the Lyapunov stability constraint in (\ref{eq:thm:2}), the following QP can be written to meet the stability and safety objectives: 
\begin{align}\label{eq:ZBFLF-QP}
\left[W^{*^{T}},\ \delta^{*}\right]^{T} & =\arg\min_{\left(W,\ \delta\right)\ \in\ \mathbb{R}^{n_{h}+1}}\:E_{D}+p\delta^{2}\\
\mathrm{s.t.}\qquad &  -\nabla h\left([x]_{\tau}\right)^{T}W^{T}g-\gamma h([x]_{\tau})\leq-\mathcal{E},\nonumber \\
 & \left([x]_{\tau}-x^{*}\right)^{T}W^{T}g\leq-\beta\left([x]_{\tau}\right)-\mathcal{L}_{V}\frac{\tau}{2}+\delta, \nonumber
\end{align}
for all $[x]_{\tau}\in\mathcal{X}_{\tau}$, and $p$ is a positive constant. 
An auxiliary relaxation variable $\delta$ is introduced in the Lyapunov constraint (\ref{eq:thm:2}) to soften the stability objective which maintains the feasibility of the QP. 

\begin{remark}
An unconstrained ELM parameter learning scheme can be used to obtain upper bounds on $\epsilon\left([x]_{\tau}\right)$ and $\epsilon^{\prime}\left([x]_{\tau}\right)$ to be used in $\mathcal{E}$ and $\mathcal{L}_V$. 
\end{remark}

\subsection{Robustness Analysis}
In this section, robustness analysis of the learned nonlinear dynamical system model with respect an external disturbance input is presented. 
Consider the perturbed system of (\ref{eq:compact_NN})
\begin{align}
    \dot{x}=f\left(x\right)+d\left(x\right), \label{eq:perturbed}
\end{align}where $d(x)$ is a bounded disturbance, i.e., $\Vert d(x) \Vert_{\mathcal{L}_{\infty}} \leq \bar{d}$. 

\begin{lemma}\label{lem:robustness}
Given Assumptions \ref{assump_1}-\ref{assump_3} hold and the optimization problem in (\ref{eq:ZBFLF-QP}) has a feasible solution, the learned model is finite gain $\mathcal{L}_\infty$ stable with respect to the external disturbances $d(x)$ such that for a bounded disturbance $\Vert d(x) \Vert_{\mathcal{L}_{\infty}}\leq \bar{d}$ and $x(t)\in \mathcal{X} \backslash \mathcal{X}_d$, where $\mathcal{X}_{d} = \{x\in\mathbb{R}^{2}\!:\! h(x)\geq-\kappa(\Vert d(x)\Vert_{\mathcal{L}_{\infty}})\}$ and $\kappa(\cdot)$ is a class $\mathcal{K}$ function, the following inequality holds  $\underset{t\rightarrow\infty}{\lim\sup} \Vert x-x^* \Vert \leq \frac{\bar{\epsilon} + \bar{d}}{\rho}$.
\end{lemma}

\begin{IEEEproof}
The proof is given in two steps. First the robustness result of ZBF under non-vanishing external disturbances acting on the system is used \cite{xu2015robustness} to prove that the trajectories of system (\ref{eq:perturbed}) will asymptotically converge to an invariant set whose size depends on the magnitude of the disturbance. Then a Lyapunov analysis is used to prove that the trajectories of the system (\ref{eq:perturbed}) will converge to a bound around the equilibrium inside the the invariant set and robustness is shown in the sense of $\mathcal{L}_\infty$. 

For the first step of the proof, let $\mathcal{X}$ be an open set on which a continuously differentiable barrier function $h(x):\mathcal{X}\rightarrow \mathbb{R}$ is defined for the nonlinear dynamical system in (\ref{eq:compact_NN}). Let $\mathcal{X}_0$ be a closed set defined by $h(x)$ and $\mathcal{X}_{d} = \{x\in\mathbb{R}^{n}\!:\! h(x)\geq-\kappa(\Vert d(x)\Vert_{\mathcal{L}_{\infty}})\}$, where $\kappa(\cdot)$ is a class $\mathcal{K}$ function, be a closed set such that $\mathcal{X}_0 \subset \mathcal{X}_d$, and $\mathcal{X}_{d} \subseteq \mathcal{X}$. Using the Proposition 1 of \cite{xu2015robustness} the set $\mathcal{X}_0$ is asymptotically stable (see Definition \ref{def:2}). Hence, for the set $\mathcal{X}_{d}$, if any disturbances bounded by $\bar{d}$ pushes the state into $\mathcal{X}\backslash \mathcal{X}_{d}$, the set  $\mathcal{X}_{d}$ is asymptotically reached, proving the robustness of the barrier constraint.

 To prove the robustness of the Lyapunov constraint with respect to an equilibrium, let us consider the Lyapunov candidate function $V(x)$ selected in the proof of Lemma \ref{lem:UUB}. Differentiating it along the solution of (\ref{eq:perturbed}) gives $\dot{V}=\left(x-x^{*}\right)^{T}\left(W^{T}g+\epsilon(x)+d(x)\right)$. To facilitate the analysis, let us define the variable $u\left(x\left(t\right)\right)=\epsilon(x\left(t\right))+d(x\left(t\right))$ and the function $f\left(x,u\right)=W^{T}g+u$ for all $\left(x,u\right)\in\mathcal{X}\times D_{u}$ such that $\left\{ \Vert u\Vert\leq r_{u}\right\} \subset D_{u}\subset \mathbb{R}^2$ with $r_{u}>0$. From the definition we have $\Vert f\left(x,u\right)-f\left(x,0\right)\Vert\leq\Vert u\Vert$. Using the definition of the derivative of the Lyapunov function, the constraint in (\ref{eq:LF_constr}) can be rewritten as follows
\begin{align}
\dot{V}(x)\leq-\rho\Vert x-x^{*}\Vert^{2}+\Vert x-x^{*}\Vert\Vert u\Vert
\end{align}
Using the comparison Lemma and given the choice of the Lyapunov function, the euclidean norm of the error is given by
\begin{align}
    \Vert x(t)\!-\!x^* \Vert \leq \Vert x(t_0) - x^* \Vert e^{\frac{-\rho t}{2}} \! + \! \frac{1}{2}\int_{0}^{t} e^{\frac{-\rho (t-\omega)}{2}} \Vert u(\omega)\Vert d\omega \label{eq:euc_err_norm} 
\end{align}
Using the result of Theorem 5.1 of \cite{Khalil2002} for any $x\left(t_{0}\right)$ such that $\Vert x\left(t_{0}\right)-x^{*}\Vert\leq r$ with $r>0$, for $u\left(x\left(t\right)\right)\in\mathcal{L}_{\infty e}$ with $\sup_{0\leq t\leq\Theta}\left(\Vert u\left(x\left(t\right)\right)\Vert\right)\leq\sup_{0\leq t\leq\Theta}\left(\Vert\epsilon\left(x\left(t\right)\right)\Vert+\Vert d\left(x\left(t\right)\right)\Vert\right)\leq\mathrm{min}\left\{ r_{u},\rho r\right\}$ , and for $\Theta\in[0,\infty)$ the error satisfies
\begin{align}
\Vert x\left(\Theta\right)-x^{*}\Vert_{\mathcal{L}_{\infty}}&\leq\frac{1}{\rho}\left(\Vert\epsilon\left(x\left(\Theta\right)\right)\Vert_{\mathcal{L}_{\infty}}+\Vert d\left(x\left(\Theta\right)\right)\Vert_{\mathcal{L}_{\infty}}\right) \nonumber \\  &\quad+\Vert x\left(t_{0}\right)-x^{*}\Vert\label{eq:upperbnd_LFconstr_robustness}
\end{align}
which concludes that the system in (\ref{eq:perturbed}) is small-signal finite-gain $\mathcal{L}_{\infty}$ stable with $\mathcal{L}_{\infty}$ gain less than or equal to $\frac{1}{\rho}$. Furthermore, as $t\rightarrow \infty$ in (\ref{eq:euc_err_norm}) the euclidean norm of the error is bounded by $\frac{\bar{\epsilon}+\bar{d}}{\rho}$, i.e., $\underset{t\rightarrow\infty}{\lim\sup} \Vert x-x^* \Vert \leq \frac{\bar{\epsilon} + \bar{d}}{\rho}$.

\end{IEEEproof}

\section{Numerical Evaluations}\label{sec: results}
The proposed method is tested on the shapes from the LASA human handwriting dataset \cite{lemme2015open}. The dataset consists of $7$ demonstrations where the handwriting motions were collected from a pen input using a Tablet PC. The optimization problem is solved using the CVX package in MATLAB $2020b$. Random initialization process in the ELM algorithm may lead to having saturated or constant neurons, which are not desired while learning the model \cite{neumann2013optimizing}. To circumvent this problem, an intrinsic plasticity (IP) learning rule can be used \cite{triesch2005gradient}. IP is an online learning method that optimizes the information transmission of a single neuron by adapting the slopes and biases such that the output of the activation function becomes exponentially distributed. The estimates of the slopes and biases are obtained using a computationally efficient batch version of the IP method, called BIP presented in \cite{neumann2013optimizing}.

\begin{table*}[tbp]
\small
 \caption{The mean value of the model empirical bounds and success rate of staying within the predefined bounds calculated over $100$ Monte Carlo runs}
  \label{tab:1}
\centering
\begin{tabular}{|l|c|c|c|c|c|c|}
\hline 
\textbf{Shape} & $\mathbf{\mu_{UB}}[mm]$  & $\mathbf{\mu_{\lim}}[mm]$ & \multicolumn{3}{c|}{\textbf{Parameters}} & \textbf{Prob. of Success}\tabularnewline
\hline 
\hline 
Khamesh & $2.30$ & $1.69$ & $\rho=5$ & $\mu_{W}=0.03$ & $p=10^{-4}$ & $90\%$\tabularnewline
\hline 
Leaf-2 & $4.89$ & $4.13$ & $\rho=4$ & $\mu_{W}=0.03$ & $p=10^{-2}$ & $92\%$\tabularnewline
\hline 
NShape & $2.42$ & $1.79$ & $\rho=7$ & $\mu_{W}=10^{-9}$ & $p=10^{-9}$ & $89\%$\tabularnewline
\hline 
RShape & $2.58$ & $2.22$ & $\rho=5$ & $\mu_{W}=0.01$ & $p=10^{-3}$ & $94\%$\tabularnewline
\hline 
Multi-Models-2 & $8.79$ & $8.41$ & $\rho=3$ & $\mu_{W}=0.01$ & $p=10^{-9}$ & $90\%$\tabularnewline
\hline 
\end{tabular}
\end{table*}

The objective is to learn the parameters of the ELM subject to safety and stability constraints given by zeroing barrier function and Lyapunov stability constraints in order to guarantee that the solutions of the learned system model stay within a predefined closed set and are stable with respect to an equilibrium point inside the closed set. The selected closed set encloses all the trajectory data used for learning. Two shapes are chosen for the safe regions, an ellipse and a circle. 

\begin{figure}[tbh]
  \centering
    \includegraphics[width=0.95\columnwidth]{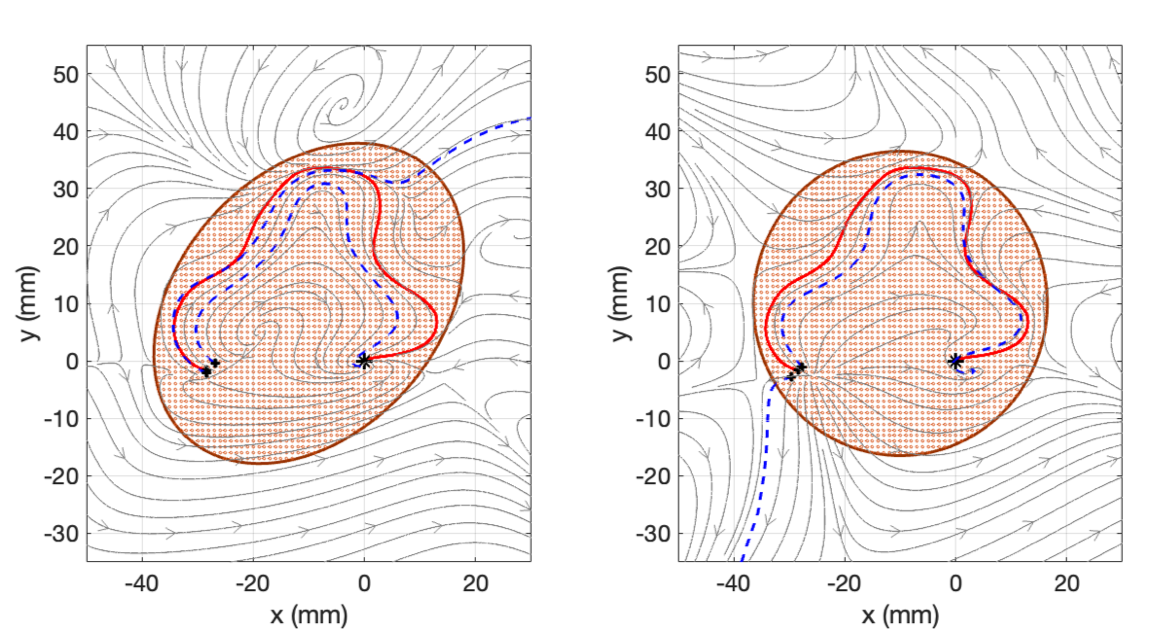}
    \caption{Model \textit{Leaf-2}, learned only with Lyapunov stability constraints for ellipse and circle safe sets.\label{fig:LF-only}}
\end{figure}

\begin{figure}
   \centering
    \includegraphics[width=0.95\columnwidth]{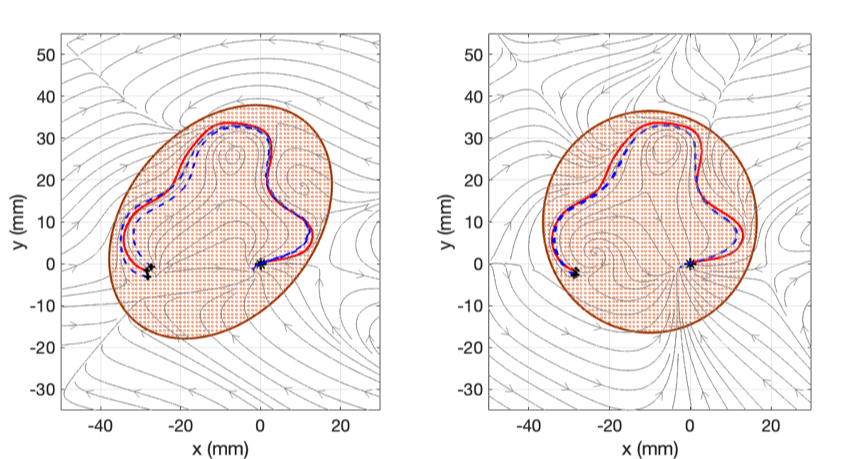}
    \caption{Model \textit{Leaf-2}, learned with zeroing barrier and Lyapunov stability constraints for ellipse and circle safe sets.\label{fig:ZBFLF}}
\end{figure}

Once the model is learned, it is used to forward propagate the learned dynamics from various initial conditions inside the invariant region. Fig. \ref{fig:LF-only} illustrates reproduction of trajectories using a learned model that uses only stability constraints. The streamlines of the solution trajectories (in gray) are shown along with the demonstrated data in solid red lines and the reproductions are shown in dashed blue lines. As seen in Fig \ref{fig:LF-only}, some trajectories that are starting inside the safe set transgress the safe region. On the other hand, the trajectories of the models that are learned with both safety and stability constraints, shown in Fig. \ref{fig:ZBFLF}, remain inside the predefined safe set for all time and converge to a bound close to the equilibrium point.

\begin{figure}[tbh]
   \centering
    \includegraphics[width=1\columnwidth]{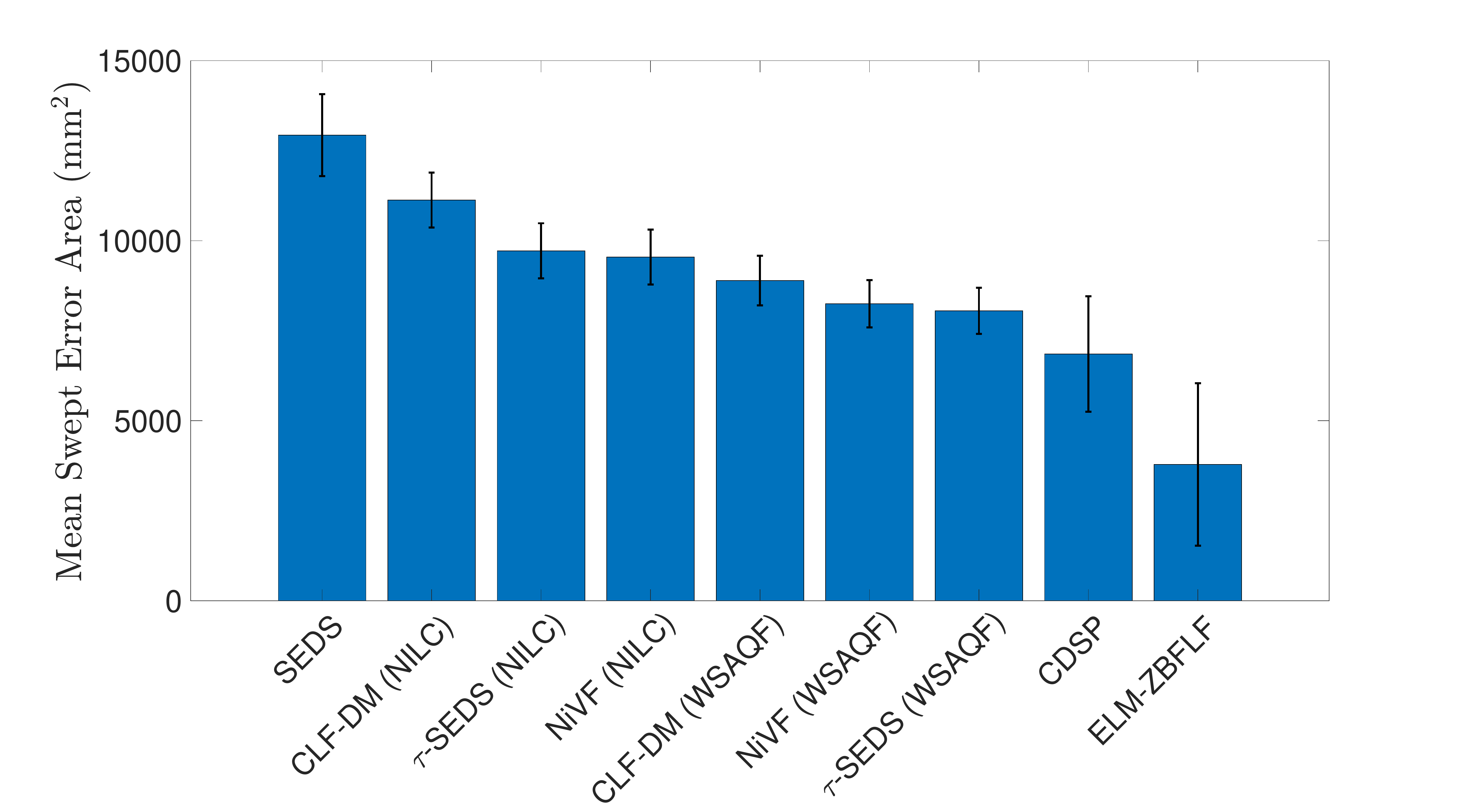}
    \caption{The SEA of different algorithms computed on the library of $30$ handwritten motions each with $7$ different demonstrations.\label{fig:SEA}}
\end{figure}

To evaluate the motion reproduction accuracy swept error area (SEA) \cite{khansari2014learning} given by
\begin{align}
    \mathrm{SEA}\!=\!\frac{1}{N_d}\sum_{n=1}^{N_d}\sum_{t=0}^{T_{n} - 1}\mathcal{A}\left(\hat{x}_n(t),\hat{x}_n(t+1),x_n(t),x_n(t+1)\right), \label{eq:SEA}
\end{align}
is used, where $\hat{x}_n(t), \ \forall t = 0 ,\cdots,T_n$ is the equidistantly re-sampled reproduction of the $n$th demonstration with $T_n$ samples, $N_d$ is the number of demonstrations, and $\mathcal{A}(\cdot)$ denotes the area of enclosed tetrahedron formed with the points $\hat{x}_n(t),\ \hat{x}_n(t+1),\ x_n(t)$, and $x_n(t+1)$ as corners \cite{khansari2014learning}. The SEA metric of the ELM-ZBFLF method is compared against the following state-of-the-art LfD methods: stable estimator of dynamical systems (SEDS) \cite{khansari2011learning}, control Lyapunov function-based dynamic movement (CLF-DM) using neurally imprinted Lyapunov candidate (NILC) and weighted sum of quadratic functions (WSAQF) methods to parameterize the energy function \cite{khansari2014learning}, neurally imprinted vector fields (NiVF) \cite{lemme2013neurally} using NILC and WSAQP methods, $\tau$-SEDS \cite{neumann2015learning}, and contracting dynamical system primitive (CDSP) \cite{chaandar2019learning}. The means and standard deviations of SEA for all the state-of-the art algorithms used in this comparisons are acquired from \cite{chaandar2019learning}. The result of the comparisons of means and standard deviations of SEA are summarized in Fig. \ref{fig:SEA}.

Based on the complexity of the shapes in LASA dataset, the results presented in Fig. \ref{fig:SEA} are computed using an ELM network with hidden neurons ranging from $n_h=25$ to $n_h=100$ where the regularization parameters in the optimization problem in (\ref{eq:ZBFLF-QP}) decrease from $\mu_W = 10^{-2}$ and $p=10^{-2}$ to $\mu_W = 10^{-9}$ and $p=10^{-9}$. The upper bounds $L_f=0.01$ and $L_V=0.01$ are selected empirically, and $\bar{\epsilon}$ as well as $\bar{\epsilon}^{\prime}$ are obtained using unconstrained ELM parameter learning scheme. Other parameter values are selected as follows $\gamma=2$ and $\tau = 10^{-9}$.

The results in Fig. \ref{fig:SEA} reveal that the average SEA of ELM-ZBFLF is significantly smaller than the other approaches. This is due to the fact that the selection of the invariant region for each shape is made separately only to cover the $7$ demonstrations. Therefore, the reproduced trajectories always stay within the invariant region. In addition, the Lyapunov constraint enforces exponential convergence to the demonstrated trajectories within an ultimate bound which can be reduced by adjusting the parameters of the constraint.
\begin{figure*}
\centering
    \includegraphics[width=1\textwidth]{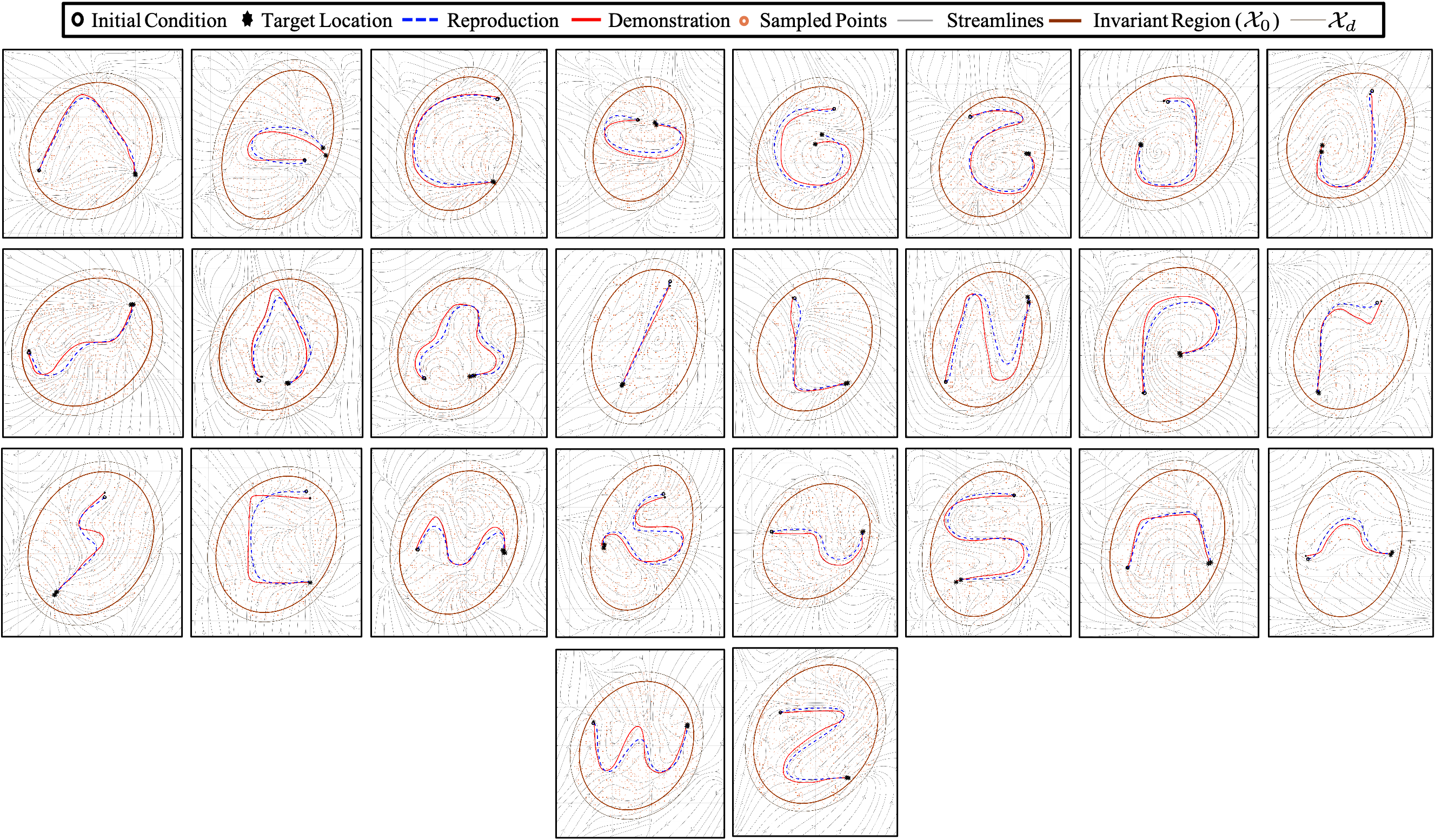}
    \caption{Qualitative performance of ELM-ZBFLF approach on LASA dataset with a normally distributed Gaussian noise. For the sake of clarity, only $1$ of $7$ demonstrations is shown. \label{fig:LASA_ZBFLF}}
\end{figure*}
To illustrate the robustness of the learned model with both zeroing BF and Lyapunov stability constraints, a normally distributed Gaussian noise with mean $\mu=2$ and variance $\sigma = 2$ that models the unpredictable disturbances for the nonlinear system dynamics (\ref{eq:perturbed}) is considered. The bound derived in Lemma \ref{lem:robustness}, i.e., $\underset{t\rightarrow\infty}{\lim\sup}\Vert x-x^* \Vert \leq \frac{\bar{\epsilon} + \bar{d}}{\rho}$, is used to check for violation over $100$ Monte Carlo runs. For brevity, Table \ref{tab:1} summarizes the results of only $5$ out $10$ shapes. In Table \ref{tab:1}, $\mu_{UB}$ denotes the mean bounds, which is given by $\frac{1}{100}\sum_{i=1}^{100}\frac{\bar{\epsilon}(i)+\bar{d}(i)}{\rho}$, and $\mu_{\lim}=\frac{1}{100}\sum_{i=1}^{100}\max_{k\geq\bar{k}}\Vert x_{i}\left(k\right)-x^{*}\Vert$ where $x_{i}\left(k\right)$ is the $k^{\mathrm{th}}$ trajectory point in the $i^{\mathrm{th}}$ Monte-Carlo run and $\bar{k}=991$ for a total of $1000$ trajectory points. Fig. \ref{fig:LASA_ZBFLF} illustrates the qualitative performance of the proposed method on the LASA dataset when a normally distributed Gaussian noise with the above mentioned mean and variance values is added during forward integration of the learned model. Note that all the $7$ demonstrations for each shape are used to train the model, but for the sake of clarity only one of them is shown in Fig. \ref{fig:LASA_ZBFLF}; also a total of $1000$ points are randomly sampled from set $\mathcal{X}_d$ to train the model, and orange dots show them in the figure.

\begin{figure}
   \centering
    \includegraphics[width=1\columnwidth]{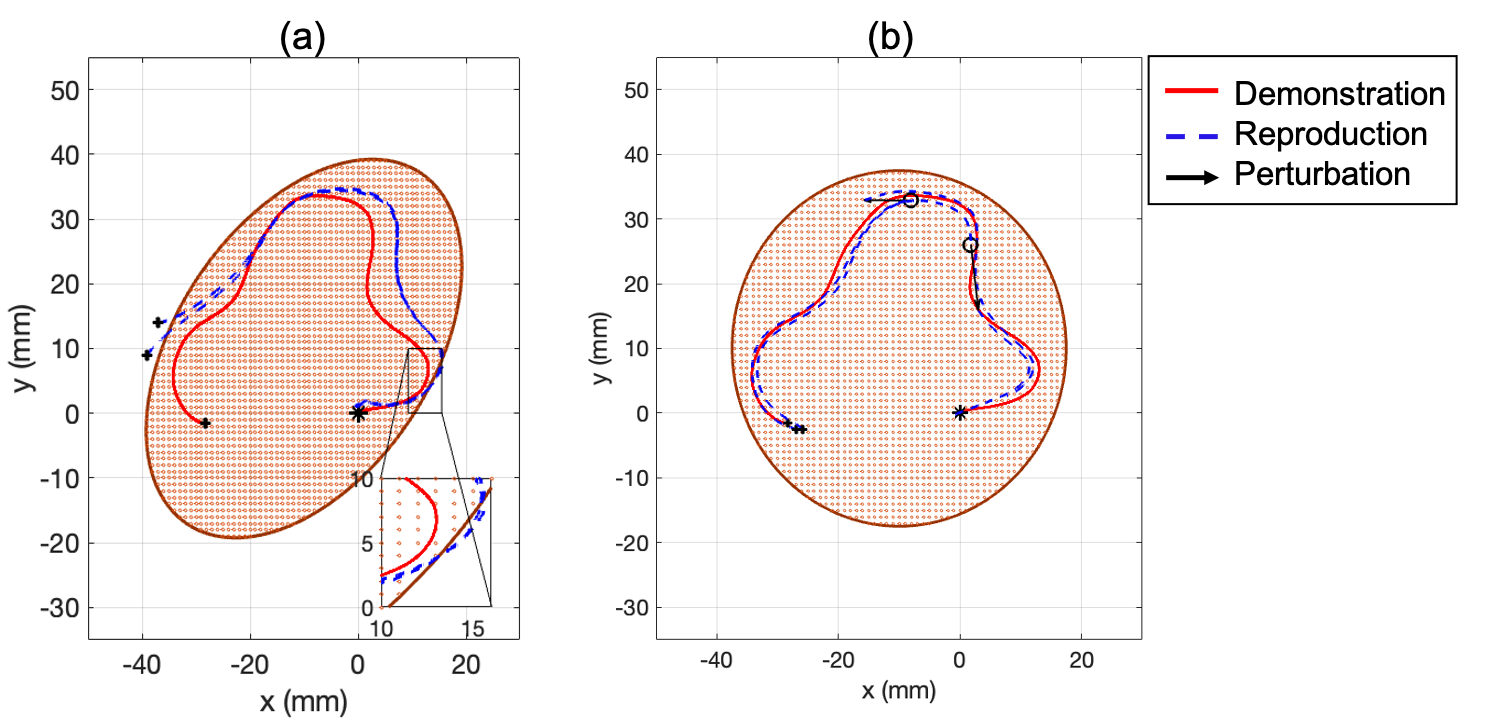}
    \caption{Illustration of the model's ability to handle external disturbance and perturbation when it is learned with zeroing barrier function and Lyapunov stability constraints.\label{fig:ZBFLF_perturb}}
\end{figure}

Fig. \ref{fig:ZBFLF_perturb}-a presents a scenario that the safe set is selected very tightly around the demonstrations. It can be seen that during the trajectory reproduction and within the presence of a Gaussian disturbance with mean $\mu = 2 \ [\mathrm{mm}]$ and variance $\sigma^2 = 2 \ [\mathrm{mm}]$, the states are pushed outside the safe set. Still, they come back inside and remain within the corresponding bound of the equilibrium point. Moreover, a discrete perturbation is also applied to the position updates at random time instance $t$ with amplitude $\mathbf{a}$ and direction $\mathbf{v}$ according to the systemic method described in \cite{lemme2015open}. Fig. \ref{fig:ZBFLF_perturb}-b shows the result of the disturbances applied at random time instances and corresponding trajectory reproduction using learned constrained ELM model.

\begin{figure*}
\includegraphics[width=1\textwidth]{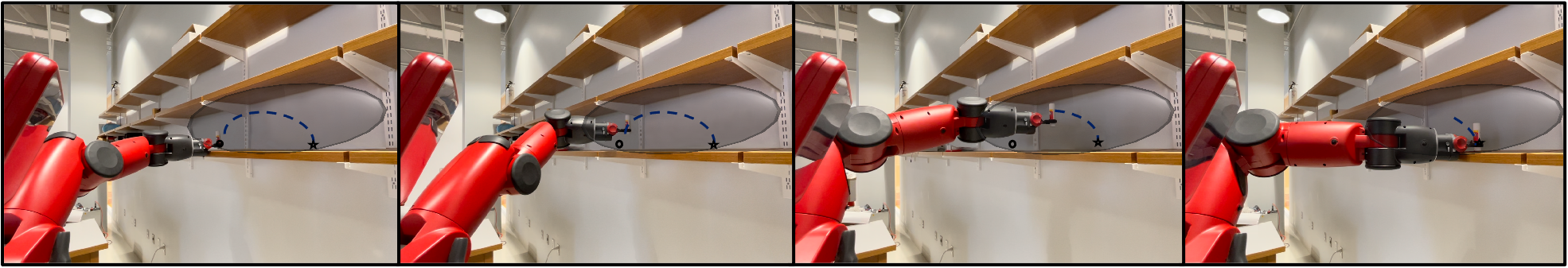}

\caption{Sequence of images show Baxter robot recreating the desired trajectory
without hitting the top and bottom shelves.\label{fig:baxter_img}}
\end{figure*}
\section{Robot Implementation}

To motivate and showcase the usefulness of the proposed method in a practical example, consider NASA's environmental control and life support system (ECLSS), a regenerative life support system that provides clean air and water to the international space station. The system consists of two main components water recovery system and an oxygen generation system that is stored inside what is known as the Universal ECLSS pallet, which is a very compact space \cite{stapleton2016environmental}. Automating the water filter replacement task inside the pallet is an activity that requires robot motion planning algorithms so that they can manipulate objects in a confined space without damaging the nearby equipment. The experiment chosen to demonstrate the utility of the proposed method for robot motion generation emulates such a scenario wherein a grasped object needs to be manipulated inside a confined space.

The task chosen for the experiment is `pick-and-place', wherein Baxter, the seven degree-of-freedom research robot, is required to pick up an object from a location on a shelf and place it elsewhere on the same shelf without colliding with the shelf. A set of $5$ demonstrations is collected by guiding the robot arm through the desired motion, and the end-effector positions and velocities are recorded. The ELM is trained on the recorded position and velocity trajectories using the safety and stability constraints described in the above sections. The built-in low-level controller and inverse kinematics engine, IKFast, are used to convert the demonstrations from Cartesian space to joint space suitable to execute the reproduced trajectories.

\begin{figure}[tbh]
\centering
\includegraphics[width=1\columnwidth]{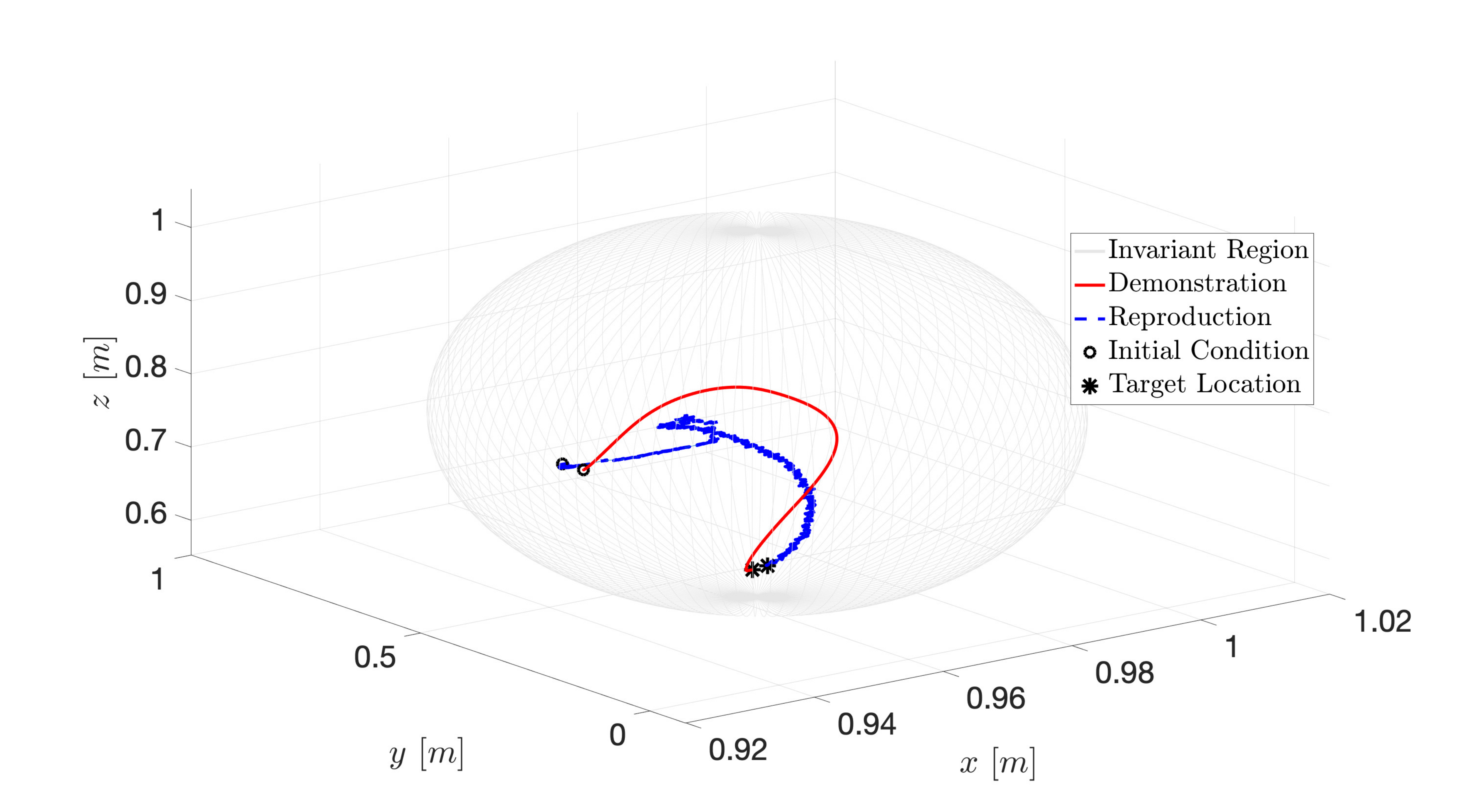}

\caption{Baxter implementation results showing reproductions using a model trained with stability and safety constraints. For the sake of clarity, only $1$ of $5$ demonstrations is shown. \label{fig:baxter_rep}}
\end{figure}

Fig. \ref{fig:baxter_img} is an example of the experimental setup, which shows the robot's motion along with the reproduced waypoints without hitting the top shelf or even exiting the area that is safe for the robot to operate. Without  loss  of  generality  and  consistent  with  our  problem  formulation,  an ellipsoid is chosen to represent the critical region that the robots are allowed to operate. As it can be seen in Fig. \ref{fig:baxter_rep}, the underlying nonlinear dynamical system that governs the motion is learned such that the end-effector generated trajectories are kept inside the confined space, i.e., the ellipsoid, and they converge very close to the vicinity of the desired target location. 

\section{Conclusion}

A learning method using the ELM algorithm with zeroing barrier and Lyapunov constraints is presented. To ensure the trajectories generated by the learned model remain inside a safe set, and converge to the desired goal given by the system equilibrium, safety and stability constraints are enforced while learning the ELM model parameters. Safety is guaranteed by verifying the forward invariance of a set with respect to the system model using a zeroing BF. The stability constraints are obtained using the Lyapunov analysis. The numerical results shown in Section \ref{sec: results} reveal that the constrained ELM parameter learning guarantees that the solutions of the learned system model remain inside the circle and ellipse and converge to the system equilibrium. Experiments conducted on Baxter robot show the feasibility of the method's implementation on a real robot.

\bibliographystyle{IEEEtran}
\bibliography{TCST}

\end{document}